# Microwave Photonics Parallel Quantum Key Distribution

José Mora, Antonio Ruiz-Alba, Waldimar Amaya, Alfonso Martínez, Víctor García-Muñoz, David Calvo and José Capmany
Optical and Quantum Comunications group, ITEAM Research Institute, Universitat Politécnica de Valencia, Spain

**The incorporation of multiplexing techniques used in Microwave Photonics to Quantum Key Distribution (QKD) systems bring important advantages enabling the simultaneous and parallel delivery of multiple keys between a central station and different end-users in the context of multipoint access and metropolitan networks, or by providing higher key distribution rates in point to point links by suitably linking the parallel distributed keys. It also allows the coexistence of classical information and quantum key distribution channels over a single optical fibre infrastructure. Here we show, for the first time to our knowledge, the successful operation of a two domain (subcarrier and wavelength division) multiplexed strong reference BB84 quantum key distribution system. A four independent channel QKD system featuring 10 kb/s/channel over an 11 km link with Quantum Bit Error Rate (QBER) < 2 % is reported. These results open the way for multi-quantum key distribution over optical fiber networks.**

Microwave Photonics (MWP)[1,2], the science and engineering field dealing with the study of photonic devices operating at microwave frequencies, and their application to microwave and optical systems is now expanding to address a number of novel and emerging applications, such as optical packet switching, optical probing, terahertz-wave generation and processing for non-invasive high resolution sensing and quantum communications. Within this last area, one of the most important application is Quantum Key distribution (QKD)[3,4], in which techniques that rely on exploiting the laws of quantum mechanics are developed with the objective of sharing a random sequence of bits between two users, Alice and Bob, with a certifiably security not attainable with either public or secret-key classical cryptographic systems. Photonics has proved to be one of the principal enabling technologies for long-distance QKD using optical fiber links and several techniques have been proposed in the literature[5-18]. Initially investigated for point-to-point links, there is an increasing interest in its extension to network environments[19-22] where the use of multiplexing techniques can bring an added value for multiuser operation. The first reported results[20,22,23], based on wavelength division multiplexing (WDM) have explored the impact of one or several classical information channels over a solitary QKD channel, usually placed in a different spectral band, identifying the spontaneous Raman scattering as the dominant impairment from the strong signals. Very recently, a first WDM based QKD system using three different wavelengths in the C-Band has been reported[24,25], featuring promising results which include a 200 kb/s key generation rate with a 14.5 dB transmission loss using 1.22 GHz pulse generation rate. WDM multiplexing alone however, has the drawback of consuming a full wavelength channel for each key. A particularly interesting approach to distribute more than one key per wavelength is Subcarrier Multiplexed Quantum Key Distribution[26] (SCM-QKD), a technique borrowed from MWP, which brings several advantages such as high spectral efficiency compatible with the actual key rates achieved by QKD systems, the sharing of the optical source by all the multiplexed channels, which reduces the complexity of the system and the possibility of upgrading with WDM in a two-tier scheme, to increase the number of parallel keys and to coexist with other classical information channels over the same fiber infrastructure.

This paper provides the first-ever, experimental demonstration of both SCM-QKD and WDM/SCM-QKD systems. The first case is implemented by independently modulating two subcarriers at 10 and 15 GHz which, in turn, modulate an optical carrier. The compound signal is delivered through a dispersion compensated 11 km optical fiber length, representing a standard access network link, after which it is re-modulated and the sidebands optically filtered previous to detection. Using a source providing 1 MHz pulse repetition rate we demonstrate the generation of two independent keys at 10 kb/s featuring a quantum bit error rate (QBER) below 2 % under a total system loss of 6.5 dB. The system is then WDM upgraded with a second optical carrier to demonstrate four independent 10 kb/s keys with QBER < 2 %. In both cases, a classical reference channel used for system stabilization is also sent along the fiber link with a negligible impact due to Raman scattering.

**Results**

**Basic principles.** The operation principles of SCM-QKD[26] can be explained referring to Figure 1. A faint pulse laser source emitting at frequency $\omega_o$ is externally modulated by N radiofrequency subcarriers $\Omega_n$ (n=1, 2,… N) at Alice's location such that the mean photon number per pulse emitted by the laser source for subcarrier $\Omega_n$ verifies $\mu_n \leq 1$. For parallel key distribution, each subcarrier transmits to a given user a different key which is generated by an independent voltage controlled oscillator (VCO) randomly phase-modulated among four possible values 0, $\pi$ and $\pi/2$, $3\pi/2$ which form a pair of conjugate bases required to implement the Bennet-Brassard BB84 protocol[26,27]. As example, Figure 1a depicts the probability distribution for a situation where Alice transmits 6 keys in parallel. The compound signal is then sent by an optical fiber link through an optical network and, upon reaching Bob's location, is externally modulated by N identical subcarriers (now randomly phase-modulated among two possible values: 0 and π/2)[26,27] in a second modulator. As a consequence, an interference single-photon signal is generated at each one of the sidebands (upper and lower) of each subcarrier with a certain probability as shown in Figure 1b. For a given user, the detection probabilities at each one of the detectors placed after the filters centered at the Upper Sideband (USB) and the Lower Sideband (LSB) corresponding to $\omega_o+\Omega_n$, and $\omega_o-\Omega_n$, respectively, are given by:

$$p_{USB}(\omega_0 + \Omega_n) = \rho\mu_n T_n \cdot (1 + V\cos\Delta\Phi_n)/2$$
$$p_{LSB}(\omega_0 - \Omega_n) = \rho\mu_n T_n \cdot (1 - V\cos\Delta\Phi_n)/2$$
(1)

In the above expression, ρ is the detection efficiency, $T_n$ is the end-to-end optical link transmission efficiency for subcarrier $\Omega_n$, V is the system visibility, and $\Delta\Phi_n = \Phi_{An} - \Phi_{Bn}$ represents the mismatch between the phases $\Phi_{An}$ and $\Phi_{Bn}$ inscribed by Alice and Bob, respectively, into the subcarrier at $\Omega_n$.

For a given subcarrier $\Omega_n=2\pi f_n$, if Alice and Bob's bases match, then the photon will be detected with probability 1 by either the detector placed after the filter centered at $\omega_o+\Omega_n$, or by the detector placed after the filter centered at $\omega_o-\Omega_n$ depending on whether a "0" or a "1" is respectively encoded. If, on the contrary Bob and Alice's bases do not match there will be an equal probability of ½ of detecting the single photon at any of the two detectors and this detection will be discarded in a subsequent procedure of public discussion. Note that Figure 1b plots the measured bits at Bob's side when the base choices match in three channels and Figure 1c corresponds with the user n=3 receiving correctly the bit "0".

A further advantage of the SCM-QKD technique is that it can be combined with M WDM carriers to provide a two-tier multiplexing scheme. In this way, the overall number of parallel keys which can be distributed is given by a NM factor. Another option is to combine all the parallel keys to compose a superkey, the bit rate of which will be given by the sum of the individual keys for each single channel.

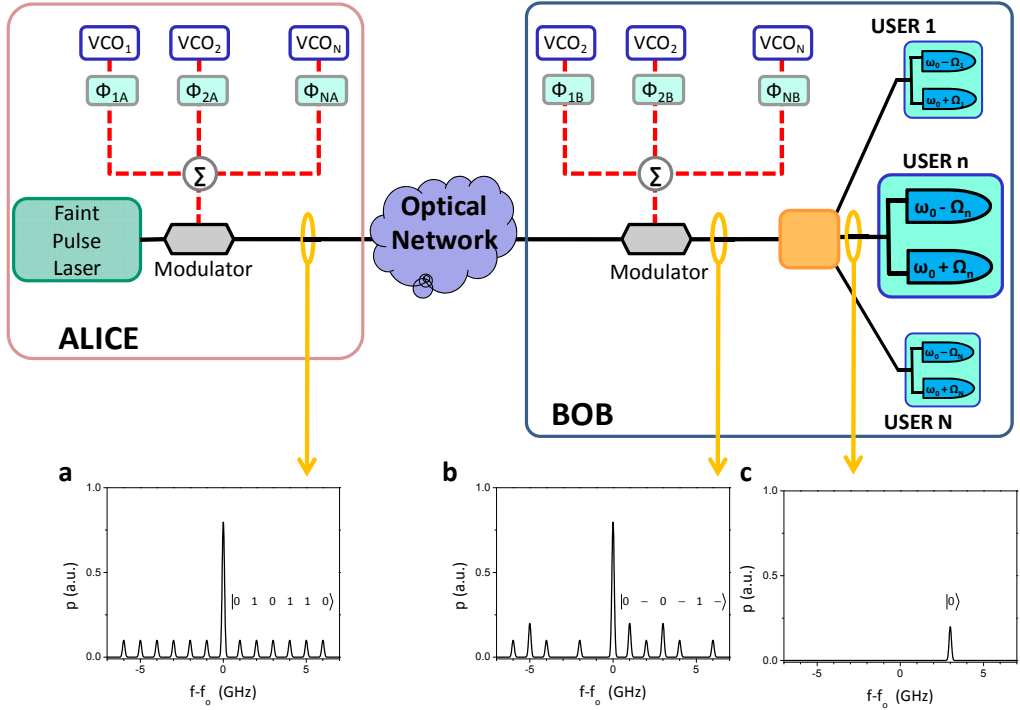

**Figure 1. SCM-QKD system layout to distribute N keys in parallel for N users.** Electrical signal is in dashed line and optical signal in solid line. Simulated probability distribution for (**a**) 6 bits transmitted in parallel by using both conjugated basis, (**b**) after detection in Bob's side measuring also in both basis and (**c**) result for a given user which receives correctly the bit "0".

**Experimental setup.** Figure 2 shows the first experimental setup assembled to demonstrate the feasibility of the SCM-QKD approach by multiplexing two independent keys. Four main blocks can be distinguished which correspond to the quantum transmitter (Alice), the quantum receiver (Bob), both interconnected by a 11 km fiber length representing a typical access network link, the classical reference channel and the overall electronic control system (see methods).

Alice's transmitter produced weak coherent-state pulses by strongly attenuating a laser source previously pulsed using a time gating electronic signal to drive a 20 dB extinction ratio electrooptic Mach-Zehnder modulator. The output pulses had 1.3 ns FWHM and a repetition rate of 1MHz. The nominal 3 dB laser linewidth was 10 MHz and the emission wavelength 1548.78 nm.

Quantum states to encode the binary secret keys were prepared at Alice's location by amplitude modulating the fainted laser pulses using a 20 GHz-bandwidth external electrooptic modulator (AM), biased at quadrature and fed through the RF port by two subcarriers, generated from independent local oscillators of frequencies $f_1$=10 and $f_2$=15 GHz. To implement the different versions of the BB84 protocols[16,27,28], each sideband must contain a mean photon number of $\mu \leq 1$ per pulse. In particular, for the case of BB84 protocol with strong reference[28,29] which was the one considered in our experiment, $\mu$ =1.

The control system enabled the pseudo random generation and independent impression of time varying phase shifts $\Phi_{1A}$ and $\Phi_{2A}$ onto each subcarrier in synchronicity with the arrival of the fainted pulses. 8-bit, digitally tunable phase shifters (500 ns switching speed) capable of providing full 360° phase shifts with a 1.4° resolution step were employed for that purpose. Electrical attenuators ($Att_{1A}$ and $Att_{2A}$) were placed at the input of both phase shifters to independently control the amplitude of the RF signal driving each subcarrier.

Bob's receiver has a similar configuration as Alice but in this case the optical signal after propagating through the fiber link is modulated by means of a 20 GHz-bandwidth Phase modulator (PM). Bob selects the basis for each subcarrier to realize the measurement of the transmitted qubit by synchronously inserting independent random phase shifts $\Phi_{1B}$ and $\Phi_{2B}$. After filtering, the photon detection was realized by placing an ID Quantique (id201) Single Photon Avalanche Detector (SPAD) for each optical sideband. The quantum efficiency and the dark count probability of the SPADs was 10 % and $10^{-5}$, respectively, which were operated using a time gate of 2.5 ns and synchronized with the faint pulse source by means of the control system.

A classical reference channel was required to convey a synchronization signal from Alice to Bob and also to stabilize the link against fiber length fluctuations[18] by providing Bob with exact replicas of the 10 and 15 GHz electrical subcarriers produced at Alice's side. The reference and quantum channels were coarse wavelength multiplexed (CWDM) to share the same optical fiber link. The CWDM multiplexer mixed two optical bands with a 20 nm wavelength separation and insertion losses of 0.5 dB. As shown in Figure 2b, the optical band centered at 1551 nm was used to transmit the quantum channels while the adjacent band at 1531 nm was used for the reference channel. After fiber transmission, a CWDM demultiplexer separated both quantum and reference channels.

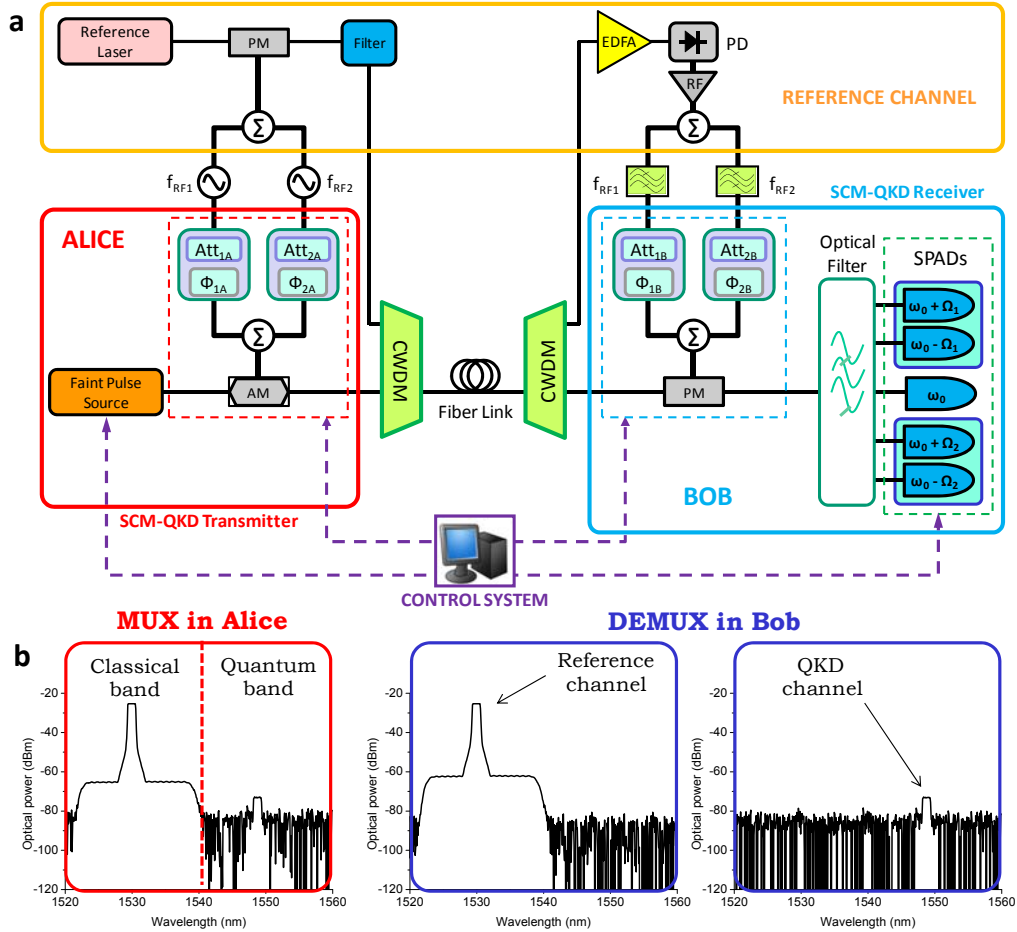

**Figure 2. a**, Experimental setup assembled in the laboratory to test the feasibility of a two keys in parallel transmission by means a SDM-QKD system (N=2). **b**, Optical spectra for classical and quantum bands when are multiplexed in Alice and Bob, respectively. Note that power level of quantum channel which would be below the noise floor is intentionally augmented to show its spectral location.

**Limiting factors.** The design of the SCM-QKD implies the solution of several challenging limitations that can drastically reduce the useful key rate: The implementation of narrowband optical filters with high extinction ratios for the selection of the RF sidebands, the control of the Raman effect due to the reference channel over the quantum band, the control of environmental fluctuations in the optical path and the compensation of chromatic dispersion.

To optically filter each one of the sidebands with enough extinction ratio at the output of Bob's modulator we designed and implemented a photonic filter structure composed of different Fiber Bragg Grating (FBG) stages[30] as shown in Figure 3a. The filter allows the extraction of the optical carrier as strong reference to guarantee unconditional security and features a high-extinction filtering of each sideband providing uniform output probability for each sideband (see Methods). We checked the filter performance in the classical regime with the aid of a 10 pm resolution ANDO optical spectrum analyzer (OSA), placed at each output of the filter. Figure 3b shows the measured spectra for all filtered bands at ±10 GHz and ±15 GHz with respect to optical carrier. All sidebands display the same power (probability) with an extinction ratio of 25 dB while introducing minimum insertion losses ($T_{FILTER}$ = 1.5 dB).

Also, we evaluated the photon crosstalk between the reference channel and the SCM-QKD channel due mainly to Raman effect[22]. Figure 3c depicts the count rate measured for each optical sideband as a function of the product of the classical power $P_{IN}$ of the reference channel and the amount of optical losses $T_B$ at Bob's receiver. The behavior is very similar among the sidebands although slight differences are found due to optical filtering nonuniformities. In practice, Bob's losses (around 4.5 dB) relax the requirements to minimize Raman photon count below the dark count for an optical input power close to -25 dBm. For this reason, the reference channel is optically amplified first after demultiplexing and electrically after detection at Bob's receiver.

The correct stabilizing operation of the reference channel was checked as shown in Figure 3d, which depicts the measurement of the QBER (see methods) over a one-hour interval, as the system needed a reset to adjust the synchronization and the optical hardware after this period. A clear difference can be observed when the reference channel is inactive (QBER close to 50 % for a long time period) as compared to when it is active. For this last case, QBER values were lower than 2 % and consequently an averaged visibility better than 96 % was achieved which are comparable with those of a back to back system (L = 0 km).

Chromatic dispersion can also degrade the system visibility and must be compensated. In our case a 1 km dispersion compensating fiber (DCF) was added to the 10 km standard fiber link. Figure 3e shows that complementarity between LSB and USB is lost for uncompensated dispersion and Figure 3f how this is recovered for both subcarriers when dispersion is compensated. Measured QBER values of around 17 % and 35 % for the 10 and 15 GHz subcarriers in the uncompensated case are reduced to values around 1.5 % for both subcarriers when the dispersion is compensated.

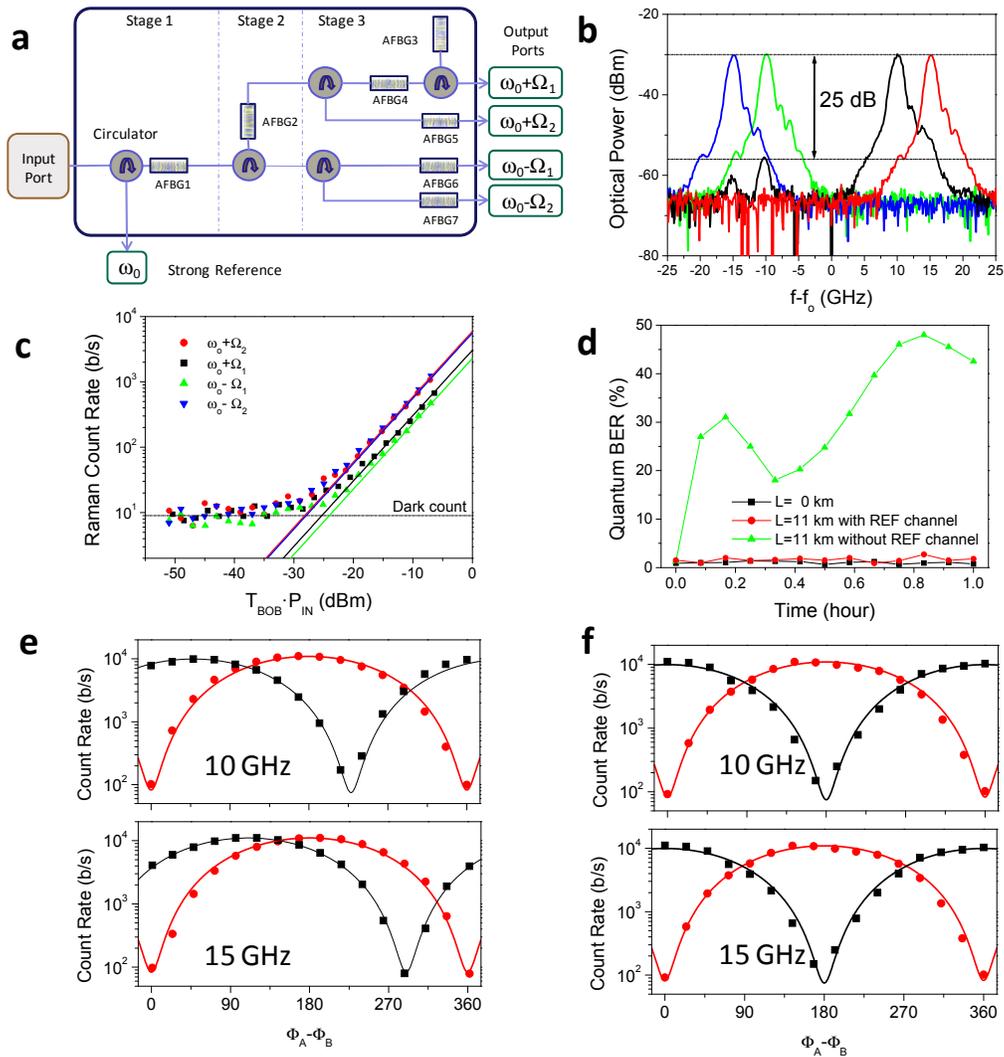

**Figure 3. a** Filter structure employed to discriminate each subcarrier. **b**, Optical spectra for each optical sideband in classical regime after the filter structure. The central frequency is gauged to the carrier $\omega_0$. **c**, Experimental count rate (points) corresponding to each optical filtering output as a function of the optical power of reference channel and Bob's losses (lines are theoretical simulations) **d**, QBER when reference channel is enable or disable on one hour for 11 km compared to back to back configuration. **e,** Count rate of sidebands for each subcarrier when the dispersion is not compensated and **f**, when is compensated. Theoretical predictions are plotted in lines.

**SCM-QKD and WDM/SCM-QKD multiplexed key and quantum bit error rates.** Performance of both SCM-QKD and WDM/SCM-QKD was tested, by measuring the individual sifted key rates and QBER (see methods), after transmission through an 11 km optical fiber link. In a first stage the feasibility of the SCM-QKD approach was proved by independently phase encoding two RF subcarriers with an averaged photon number of 1 photon per pulse at 1 MHz repetition frequency. The wavelength of the optical carrier was 1557.30 nm. Figure 4a shows the sifted key rate of the individual subcarriers (~ 10 kbit/s at 10 and 15 GHz) and the aggregated sifted key rate (~ 20 kbit/s) when both subcarriers are transmitted simultaneously. The independent key distribution using such a tightly spectral separation (5 GHz) is thus demonstrated for the first time to our knowledge. Furthermore, the aggregated sifted key rate is very close to the maximum multiplexing gain (3 dB) corresponding to the number of subcarriers N=2 as theoretically predicted[26]. The corresponding QBER is plotted in Fig. 4b featuring a value below 2 % for all count rates which implies visibilities higher than 96 % according to the signal level. As expected, this reflects the fact that the drifts due to temperature and vibrations are properly compensated by the reference channel.

Finally, to further prove the scalability and flexibility of the proposed system, a WDM/SCM-QKD system was assembled in a two-tier configuration. First, two independent SCM-QKD transmitters (each one generating two independently phase-encoded multiplexed subcarriers at 10 GHz and 15 GHz) were implemented at Alice's location centered at 1548.78 (CH1) and 1557.30 (CH2) nm, respectively. These channels were wavelength multiplexed using a Dense Wavelength Division Multiplexer (DWDM) based on FBGs and then, all quantum channels were multiplexed via CWDM with the reference channel as shown in Figure 5a. In a similar way than single SCM-QKD channel, the classical and quantum bands were recovered with a CWDM demultiplexer and each quantum channel was filtered by means of a SCM-QKD receiver after a DWDM demultiplexer. Figures 5b and 5c plot the individual sifted key rates obtained for each one of the four subcarrier channels and the aggregated sifted key rate which, in this case, is close to the M·N=4 value. Note that QBER values lower than 2 % are obtained for all the channels.

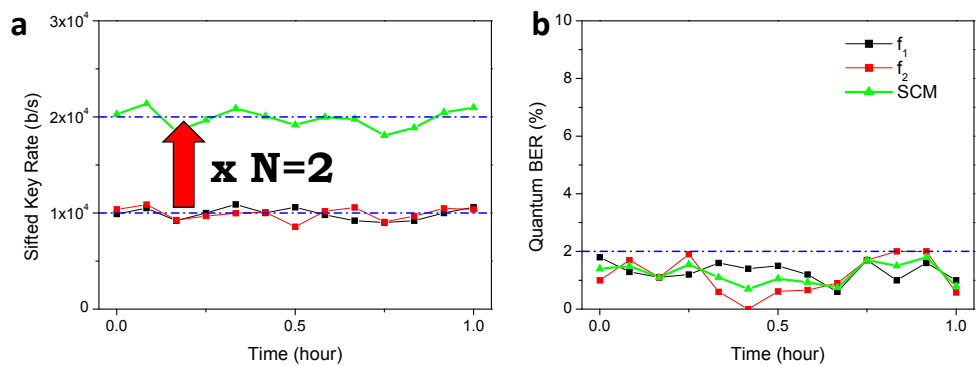

**Figure 4. a,** Evolution of the sifted key rate for each individual SCM-QKD channel and the multiplexed sifted key rate when SCM multiplexing technique is considered. **b,** Measurement of the corresponding QBER for each single channel and for the multiplexed rate.

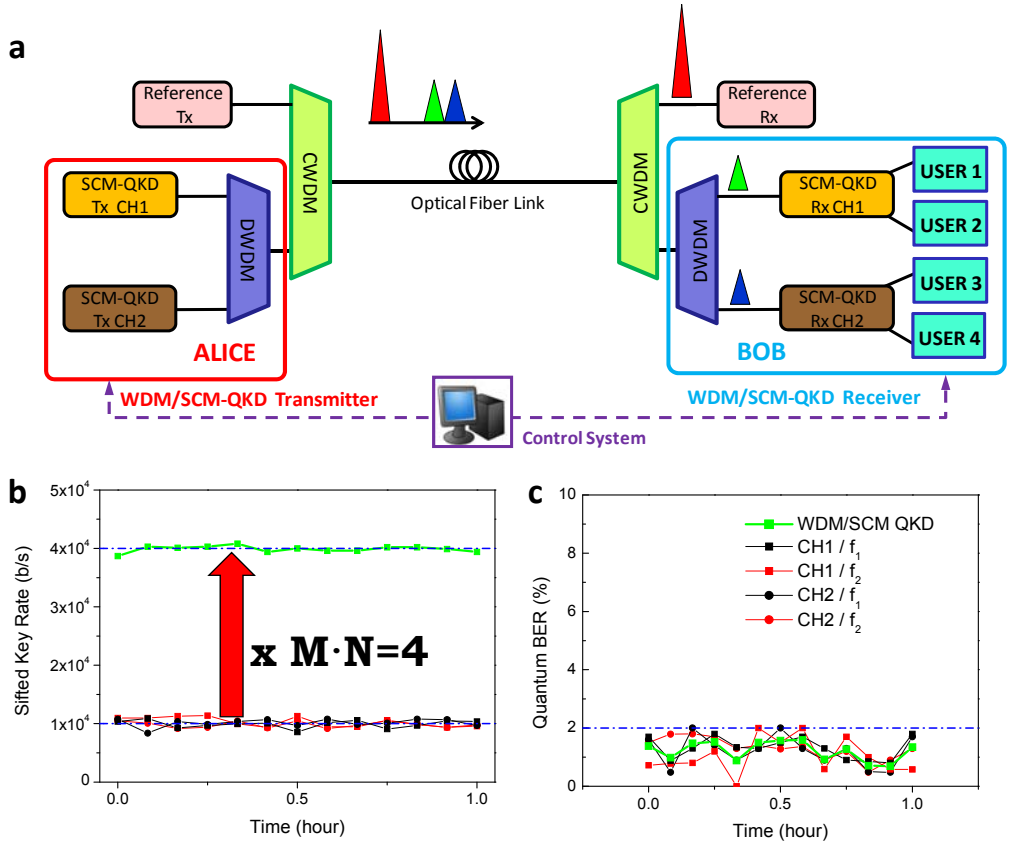

**Figure 5. a, QKD** scheme for both WDM and SCM multiplexing transmission, **b** Sifted key rate for each individual quantum channel and the multiplexed key rate. **c,** Measurement of the corresponding QBER for each single channel and for the multiplexed sifted key rate.

**Discussion**

The reported work has demonstrated that SCM, widely employed in the field of MWP, can be used in the context of QKD, to achieve the simultaneous distribution of parallel keys by using frequency channels closely packed in the optical spectrum. The importance of this advance is very significant for several reasons which are now stated.

Firstly, it relies on a technique which is very well known for its high spectral-efficiency. Since the actual key rates achieved by QKD systems are very modest in comparison with those of classical broadband communications, the use of a multiplexing technique with reduced spectral separation between adjacent channels seems a natural and sustainable choice for the delivery of multiple keys. Secondly, the optical source is shared by all the multiplexed channels which reduces the complexity of the system since all keys are carried on the same wavelength assigned in an optical network providing a reduction of system complexity, management and cost. A third advantage is that the SCM approach can be combined and upgraded with WDM to increase the number of parallel keys and to coexist with other classical information channels over the same fiber infrastructure. Again, this two tier WDM/SCM combination is borrowed from MWP.

We would like to point out that despite the fact that in the experiments reported here we considered an average number of 1 photon per pulse as required by the BB84 protocol with strong reference, the SCM-QKD system was tested for different values of the mean photon number ranging from 0.1 to 1 photon per pulse obtaining satisfactory results in all the cases with the obvious reduction in the sifted key rate. Although we experimentally demonstrated moderate single and aggregated sifted key rates, the capacity of the system can be upgraded currently in at least two orders of magnitude by using components which are commercially available as phase shifters with a switch time of 25 ns, SPADs with high-speed gating at frequencies up to 100MHz and optical filters with 32 output ports in a single device with a very narrow channel spacing (~ 5 GHz).

To summarize, we have demonstrated, for the first time, the feasibility of a quantum key distribution system based on the combination of WDM and SCM multiplexing techniques in order to further enhance the transmission capacity over optical fiber networks. The proposal permits to increase the final key rate of the quantum transmission or the distribution of parallel keys for different users. The advantage of the SCM multiplexing technique against WDM over quantum key distribution is that the same photon source is shared by N different users and consequently, the complexity of the synchronization and control system is reduced drastically when the QKD system is introduced in an optical network. The obtained results confirm that MWP is a promising technology to enhance the viability of the quantum systems.

**Methods**

**Quantum Bit Error Rate Computation, Measurement and Limiting factors.** The quantum bit error rate (QBER) defined as the ratio of wrong bits to the total number of bits received is a measure of the quality of a QKD system which takes into account the most important limiting factors. In practice, these limiting factors come from two sources; environmental instabilities which drift the system away from its ideal operation point and the wrong photon count contribution due to different crosstalk processes. Environmental changes related to the fluctuations of optical path and the state of polarization between both Alice and Bob's modulators can be controlled by means of the reference channel and thus will not be further considered for the QBER computation which in the case of the Bennet-Brassard 1984 (BB84) protocol implemented by subcarrier multiplexing is given by:

$$QBER(\Omega_n) = \frac{1}{2} \frac{(1-V_{eff})p_{signal} + p_X + d_B}{p_{signal} + p_X + d_B}$$

Where $p_{signal}$ is the probability coming from the detection of signal photons, $d_B$ is the dark count probability related to the noise source coming from photon detection and $V_{eff}$ represents the effective visibility due to the imperfections of the devices employed in the system and the fiber dispersion. $p_X$ is the crosstalk probability and takes into account the photon crosstalk contributions measured at each optical detector which contribute to a false detection. In our case, $p_X = p_{Raman} + p_{filter} + p_{imd} + p_{phn}$ which includes several sources such as Raman effect, the extinction ratio of the optical filtering, intermodulation of SCM multiplexing and the linewidth of the laser source used as quantum carrier, respectively.

In our experimental setup, the dark count minimizing the after-pulsing effects is below $10^{-5}$ and the dispersion was compensated leading to a visibility higher than 96 %. The Raman effect was reduced below the dark count for each optical sideband by properly adjusting the input power of the reference channel. The optical filtering stage has been designed to reduce the crosstalk due to adjacent subcarriers and intermodulation products (located outside of the sidebands) below 23 dB respect to signal photon probability. In order to achieve this objective, we designed an optical filtering based on apodized fibre Bragg gratings (FBG) comprising three stages (see Fig. 3a). The first one provided the strong reference by reflecting the optical carrier with a FBG of a reflection coefficient close to 99.9 %. The second stage separated the upper RF bands from the lower RF bands. Finally, the third stage filtered each one RF bands featuring over 20 dB of rejection ratio. Therefore there are 5 ports, one for each band and one more for the optical carrier. Each FBG was centered at one fixed wavelength with 1 pm of accuracy which required a temperature control system implemented placing each apodized FBG inside of individual thermal box managed by a temperature controller.

With regards to the linewidth of the optical laser used as a faint pulse source, it had to be considered for the SCM channel close to optical carrier since additional photon counts could be introduced. In our case, a moderate modulation index guarantees that a negligible crosstalk due to this source.

Finally, the measurement of the QBER was experimentally realized by comparing a sequence of bits sent from Alice to the received bits and counting the number of errors at Bob's side. According to the above expression, the predicted QBER value for any quantum channel should be around 2 % which is confirmed by all measured results.

In the manuscript, we show the sifted key rate which was obtained after basis reconciliation. The secret key rate was not obtained experimentally since Error Correction (EC) and Privacy Amplification (PA) were not implemented. However, we estimated that the secret key rate is around 31 % of the sifted key rate according to the fraction of secure bits[23] considering PNS attack and taking into account that maximum values of QBER were below 2 %.

**Control system.** The operation of the SCM-QKD and WDM/SCM-QKD systems was electronically controlled by using Virtex V (XUPV5-LX110T) Field Programmable Gate Arrays (FPGAs) from Xilinx connected to the computers via a RS-232 protocol. The electrical control signals generated by the FPGAs were distributed to the system by means of on purpose Printed Circuit Boards (PCBs). All the controlling signals were synchronized to the operating system frequency, set to 1 MHz and moreover, each one had an independently controllable delay. The main tasks of the control system are now briefly described.

At Alice's side, it provided 1.3 ns width electric pulses used to drive the faint pulse sources. In addition, the phase shifters were independently controlled for each subcarrier by means of different 8 bit signals providing the required phase shifts to implement the two maximally overlapping bases of the BB84 protocol. At Bob's side, the control system tasks included the generation of TTL type signals to trigger the photon counters in synchronicity with the arrival times of the pulses generated at Alice's location. Whenever a count was produced, the control system detected the electric signal provided by the photon counters taking into account its position on the bit stream. The control system was also in charge of storing the streams corresponding to the parallel key bits and base selections performed by Alice by an independent pseudorandom generator implemented in the corresponding FPGAs. At Bob's location, another file is stored with the streams identifying the base choices made by Bob and the count results for USB, LSB and strong Carrier Band (CB) up to a length of raw key close to 4 kbits per user, which is determined by the available memory size of the used FPGA. These files are used in the public discussion for obtaining the sifted key after basis reconciliation process for each user and the calculation of the corresponding QBER.

**References**

[1] Capmany, J. and Novak, D. Microwave Photonics combines two worlds. Nature Photonics, **1**, 319-330 (2007).

[2] Yao, J. Microwave Photonics. J. Lightwave Technol. **27**, 314-335 (2009).

[3] Gisin, N., Ribordy, G., Tittel, W., and Zbiden, H. Quantum Cryptography. Rev. Mod. Phys. **74**, 145-195 (2002).

[4] Scarani, V., Bechmann-Pasquinucci. H., Cerf. N. J., Dusek, M., Lütkenhaus, N., and Peev, M. The security of practical quantum key distribution. Rev. Mod. Phys. **81**, 1301-1350 (2009).

[5] Bennett, C.H., Bessette, F., Brassard, G., Salvail, L., and Smolin, J. Experimental quantum cryptography. J. Cryptology, **5**, 3 (1992).

[6] Muller, A., Herzog, T., Huttner, B., Tittel, W., Zbinden H., and Gisin, N. Plug and play systems for quantum cryptography. Appl. Phys. Lett. **70**, 793 (1997).

[7] Gordon, K.J., Fernandez, V., Buller, G.S., Rech, I., Cova, S.D., and Townsend, P.D. Quantum key distribution system clocked at 2 GHz. Opt. Express **13**, 3015–3020 (2005)

[8] Townsend, P.D., Rarity, J.G., and Tapster, P.R. Single-photon interference in a 10 Km long optical fiber interferometer. Electron. Lett. **29**, 634-636 (1993).

[9] Inoue, K., Waks, E., and Yamamoto, Y. Differential phase shift quantum key distribution. Phys. Rev. Lett. **89**, 037902 (2002).

[10] Takesue, H., Diamanti, E., Honjo, T., Langrock, C., Fejer, M.M., Inoue, K., and Yamamoto, Y. Differential phase shift quantum key distribution over 105 km fibre. New J. Phys., **7**, 1-12 (2005).

[11] Takesue, H., Nam, S.W., Zhang, Q., Hadfield, R.H., Honjo, T., Tamaki, K., and Yamamoto, Y. Quantum Key distribution over a 40-dB channel loss using superconducting single-photon detectors. Nature Photonics **1**, 343-348 (2007).

[12] Mérolla, J-M., Mazurenko, Y., Goedgebuer, J.P., and Rhodes, W.T. Single-photon interference in Sidebands of Phase-Modulated Light for Quantum Cryptography. Phys. Rev. Lett. **82**, 1656-1659 (1999).

[13] Mérolla, J-M., Mazurenko, Y., Goedgebuer, J.P., Porte, H., and Rhodes, W.T. Phase-modulation transmission system for quantum cryptography. Opt. Lett. **24**, 104-106 (1999).

[14] Guerreau, O., Mérolla, J-M., Soujaeff, A., Patois, F., Goedgebuer, J.P., and Malassenet, F.J. Long distance QKD transmission using single-sideband detection detection scheme with WDM synchronization. IEEE J . Sel. Top. Quantum Electron. **9**, 1533-1540 (2003).

[15] Gobby C., Yuan Z. L. and Shields A. J. Quantum key distribution over 122 km telecom fiber. Appl. Phys. Lett. **84**, 3762-3764 (2002).

[16]  Rosenberg, D., Peterson, C. G., Harrington, J. W., Rice, P. R., Dallmann, N., Tyagi, K. T., McCabe, K. P., Nam, S. , Baek, B., Hadfield, R. H., Hughes R.J. and Nordholt, J. E. Practical long-distance quantum key distribution system using decoy levels.  New J. Phys. **11**, 045009 (2009).

[17] Scheidl, T., Ursin, R., Fedrizzi, A., Ramelow, S., Ma, X.-S., Herbst, T. , Prevedel, R., Ratschbacher, L., Kofler, J., Jennewein, T., and Zeilinger, A. Feasibility of 300 km quantum key distribution. N. J. Phys. **11**(8), 085002 (2009).

[18] Stucki, D., Walenta, N., Vannel, F., Thew, R. T., Gisin, N. , Zbinden, H., Gray, S., Towery, C. R. and Ten, S. High rate, long-distance quantum key distribution over 250 km of ultra low loss fibres. N. J. Phys. **11**(7), 075003 (2009).

[19] Townsend, P.D., Phoenix, D.J.D., Blow, K.J., and Cova, S. Design of quantum cryptography Systems for passive optical Networks.  Electron. Lett., **30**, 1875-1877 (1994).

[20] Townsend, P.D. Quantum cryptography on multiuser optical fibre networks. Nature **385**, 47 – 49 (1997).

[21] Townsend, P.D.  Quantum Cryptography on Optical fiber networks. Opt. Fiber Technol., **4**, 345-370 (1998).

[22] Chapuran, T. E., Toliver, P., Peters, N. A., Jackel, J., Goodman, M. S., Runser, R. J., McNown, S.R., Dallmann, N., Hughes, R.J., McCabe, K.P., Nordholt, J.E., Peterson, C.G., Tyagi, K.T., Mercer, L., and Dardy, H. Optical networking for quantum key distribution and quantum communications. New J. Physics. **11**, 105001 (2009)

[23] Qi, B., Zhu, W., Qian, L., Lo, H-K., Feasibility of quantum key distribution through dense wavelength division multiplexing network, New Journal of Physics **12**, 103042 (2010)

[24] Tanaka, A., Tajima, A., and Tomita, A. Colourless interferometric technique for large capacity quantum key distribution systems by use of wavelength division multiplexing. Proc. 35th European Conference on Optical Communication (ECOC2009), paper 1.4.2, Vienna, Austria, (2009)

[25] Tanaka, A., Fujiwara, M., Yoshino, K., Takahashi, S., Nambu, Y., Tomita, A., Miki, S., Yamashita, T., Wang, Z., Sasaki, M., and Tajima, A. A Scalable Full Quantum Key Distribution System based on Colourless Interferometric Technique and Hardware Key Distillation. Proc. 37th European Conference and Exposition on Optical Communications, OSA Technical Digest (CD) (Optical Society of America, 2011), paper Mo.1.B.3, (2011)

[26] Ortigosa-Blanch, A., and Capmany, J., Subcarrier multiplexing optical quantum key distribution. Phys. Rev. A., **73**, (024305) (2006).

[27] Bennett, C.H., and Brassard, G. Quantum Cryptography: Public Key Distribution and Coin Tossing. Proc. IEEE International Conference on Computers Systems and Signal Processing, Bangalore, India, 175-179 (1984).

[28] Guerreau, O., Malassenet, F.J., McLaughlin, S.W., Merolla, J-M. Quantum key distribution without a single-photon source using a strong reference. IEEE Photon. Tech. Lett. **17**, 1755-1757 (2005).

[29] Capmany, J., Fernández-Pousa, C.R., Impact of Third-Order Intermodulation on the Performance of Subcarrier Multiplexed Quantum Key Distribution, IEEE J. Lightwave Technol., **29**, 3061 – 3069 (2011).

[30] Mora, J., Ruiz-Alba, A., Amaya, W., Garcia-Muñoz, V., Martínez, A., Capmany, J. Microwave photonic filtering scheme for BB84 Subcarrier Multiplexed Quantum Key Distribution. Proc. IEEE Topical Meeting on Microwave Photonics, Montreal, Canada, 286-289, (2010).

[31] Eraerds, P., Walenta, N., Legré, M., Gisin, N., Zbinden, H. Quantum key distribution and 1 Gbps data encryption over a single fibre. New J. of Phys., **12**, 063027 (2010).

**Acknowledgements:** The authors wish to acknowledge the financial support of the Spanish Ministry of Science & Innovation and the Generalitat Valenciana through projects CONSOLIDER INGENIO 2010 Quantum Information Technologies and PROMETEO GVA 2008-092 MICROWVE PHOTONICS.

**Authors' contributions:** All the authors contributed equally to the manuscript. J.C. and J.M conceived the original idea and contributed equally to design of the experimental scheme. They planned and supervised the project and J.M. guided the experiments. J.M. and A.R. performed the SCM-QKD system and designed the reference channel integrated with WDM technique. W.A. was responsible of optical filtering based on FBGs and its implementation in the WDM scheme. A.M., V.G. and D.C. developed the control system with the integration of the detection stage by V.G. The experimental measurements were performed by J.M., A.R., W.A. and D.C. while theoretical analysis was carried out by J.M. and A.R. The manuscript was written by J.M and J.C and all authors assisted substantially with discussion results.